\documentstyle[11pt,newpasp,twoside,epsf]{article}
\begin{document}
\title{Interpreting The ${\rm \bf N_{HI}}$ Power Spectrum}
\author{Anthony Minter}
\affil{NRAO, P.O. Box 2, Green Bank, WV 24944}

\begin{abstract}
In this paper I investigate what factors -- both observational
and physical -- can change the measured slope of the 
observed 21cm HI power spectrum.
The following effects can make 
the observed turbulence appear two dimensional rather than three
dimensional:  1)  if the turbulence is 
contained in a thin filament or slab; 2) if the medium has a high optical 
depth; and 3) if any method of observation or analysis is used which 
effectively limits the emission from the medium under study to a thin slab,
for example, by analyzing an individual channel map.  Straightforward analysis 
of data can give misleading or incomplete results if these effects are not 
taken into account.
\end{abstract}

\section{Introduction}

The 21cm HI line has a column density power spectrum whose slopes
are consistent with 2--Dimensional turbulence ($\alpha \sim -8/3$)
on large spatial scales 
($> 0.01$~pc) and narrow velocity ranges (Green 1993; Dickey 
\& Crovisier 1983;
Lazarian \& Stanimirovic 2001; Dickey et al. 2001).  
The slope of this power 
spectrum is closer to a 3--D, Kolmogorov--like spectrum 
($\alpha \sim -11/3$) for wider velocity 
ranges but can still be significantly different than the Kolmogorov value.
However, the electron density spectrum is consistent with Kolmogorov--like 
turbulence.  This suggests a) that electrons and neutrals have different
turbulent characteristics, or b) that there are effects which change the
measured power spectrum slopes.

Lazarian \& Pogosyan (2000) have
discussed how turbulent velocity may effect the 
observed power spectra of HI.  However, there are other factors which
Lazarian  \& Pogosyan
did not discuss which may effect the power spectra.  Among these
are opacity and filamentary structures in the HI.

\section{Theory}
The power spectrum of the observed 21cm HI emission can be written as
\begin{equation}
PS_n(\vec{k}) = C_n^2 \left( k^2 +k_o^2\right)^{-\alpha/2}
\end{equation}
where $k={2\pi \over l}$, $k_o={2\pi \over l_o}$ 
is the largest size scale of the 
turbulence and $C_n^2$ indicates the strength of the turbulence.  Assuming
that the observed HI intensity is proportional to the column density it
can be shown that 
\begin{equation}
\left< I(\vec{x}) I(\vec{x+\delta x}) \right> = C^2 \left< N(\vec{x}) N(\vec{x+\delta x}) \right>
\end{equation}
\begin{eqnarray}
\left< I(\vec{x}) I(\vec{x+\delta x}) \right> & = & C^2  \int_0^{z_o} \int_0^{z_o} \int_{-k_i}^{k_i} \int_{-k_i}^{k_i} \int_{-k_i}^{k_i} C_n^2 \left( k^2 +k_o^2\right)^{-\alpha/2} \cr
& & e^{2\pi \imath \left( k_x \delta x + k_y \delta y + k_z \left( z-z^\prime \right)\right)} dk_x dk_y dk_z dz dz^\prime
\end{eqnarray}
where $z$ and $z^\prime$ are the two lines of sight being compared.
When the observations dictate that $z_o << l_o$ then it can be shown that
\begin{equation}
PS_N(k_x,k_y) \propto \left( k_x^2 +k_y^2 + k_o^2\right)^{(1-\alpha)/2}.
\end{equation}
Thus, under common observing circumstances the column density power spectrum 
can have an index that is one less than 
the index of the density power spectrum:
$\alpha_N = \alpha_n-1$ for $z_o \in (-\infty , \infty)$
and $l \leq \vartheta(l_o)$.

\section{Models}
Models of a turbulent cloud were developed by creating a Fourier Transformed
data cube of the density.  This was done by taking the real and 
imaginary parts at a given
$k$ from a Gaussian distribution with zero mean and a 
standard deviation given by the square root of the density power spectrum
at that $k$.  This data cube was then Fourier Transformed into a real density
data cube.  A simple radiative transfer scheme was then used to convert the
data cube into the ``observed'' two dimensional HI brightness temperature. 

\subsection{Limiting the Emission Depth}
To show that the above theory is correct, slices of different thicknesses
were integrated from the same model to form the observed HI.  The observed
HI power spectra slopes as a function of thickness is shown in Figure 1.

\begin{figure}
\plotfiddle{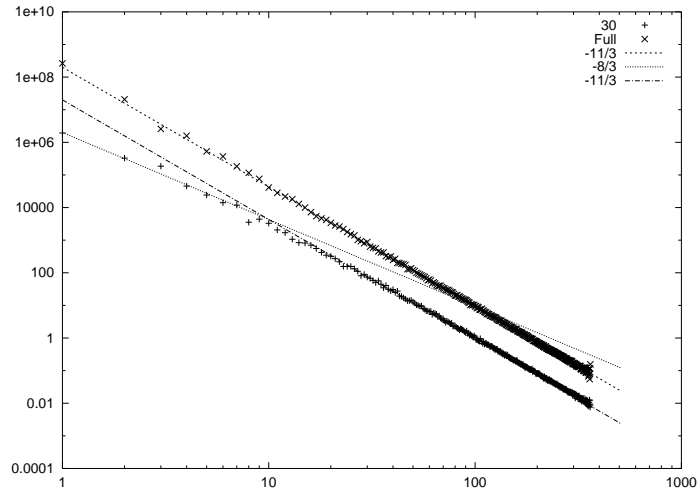}{2.5in}{0}{75}{75}{-170}{-40}
\caption{The power spectra of the same model when only the first 30 pixels in 
z are integrated over and when the full data cube are integrated over.  The
model was created with a slope of $-11/3$.  It is easily seen that any 
method of observing or a propagation effect that keeps the observer from
seeing the entire turbulent cloud will change the observed power spectrum's
slope.  The x-axis is the spacial frequency in ${\rm 512 \cdot pixels^{-1}}$ and the
y-axis is in arbitrary units.}
\end{figure}

\subsection{Opacity}
To study the effects of opacity, an average opacity from the front to the
back of the data cube was used to define the detailed radiative transfer.
As the opacity increases we can see from Figure 2 that the observed HI
power spectra change slopes from $\sim -11/3$ to $\sim -3$.

\begin{figure}
\plotfiddle{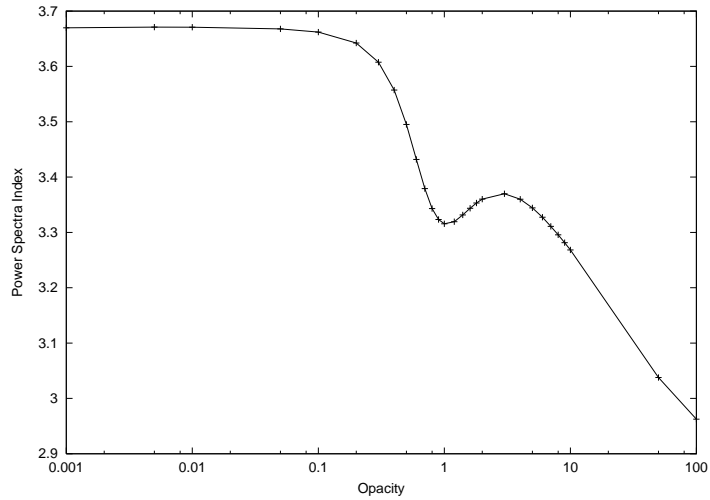}{2.5in}{0}{75}{75}{-170}{-40}
\caption{For the same model density cube this plot shows how the simulated
observed HI power spectral index varies as the opacity ($\tau$) from the
front to the back of the data cube is varied.  This plot suggests that we
may not be able to invert the observed power spectra back to the true 
power spectra since the function is not single-valued.} 
\end{figure}

\subsection{Filamentary Structures}
The HI could also be confined to thin sheets or filaments.  This view is
supported by the images of 21cm HI emission that are being produced by the
CGPS and SGPS surveys (see McClure-Griffiths 2002; and Knee 2002).
Having the turbulence confined into filaments where vortices are stretched
along the filament can also change the observed power spectral index.
This can be realized in the model by changing $k_x$ to $\eta k_x$ where
$0 < \eta \leq 1$ (i.e. making the turbulence anisotropic).  
In Figure 3 we compare the power spectra from a
an isotropic model ($\eta = 1$) to a filamentary model ($\eta=0.1$).
It is easily seen that filamentary structure can produce a power spectrum
with a slope of $-8/3$ on the largest size scales, $l > l_{filament}^\perp$
where $l_{filament}^\perp$ is the thickness of the filament.

\begin{figure}
\plotfiddle{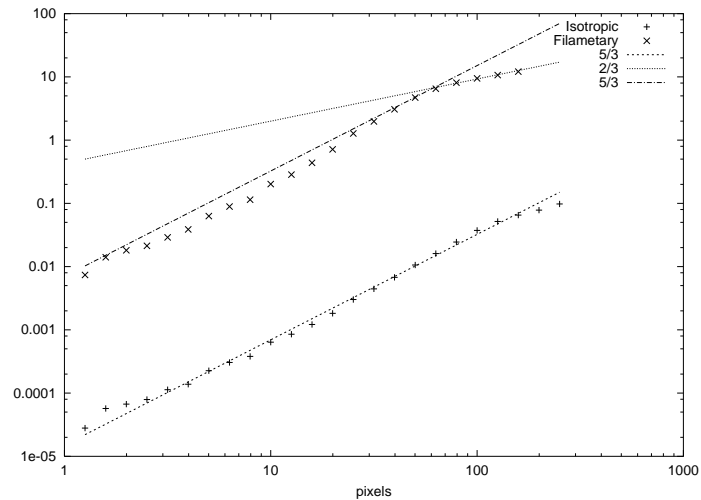}{2.5in}{0}{75}{75}{-170}{-40}
\caption{The structure functions of models of isotropic turbulence ($\eta=1$)
and turbulence in filamentary structures $(\eta=0.1$).  Adding a value of two
to the slope of the structure function and then multiplying by -1 gives the
spectral index of the corresponding power spectrum.  It is easily seen that
for the filamentary case the power spectrum slope is $-8/3$ on the largest
size scales.}
\end{figure}

\end{document}